\documentclass[11pt]{amsart}
\usepackage[left=2.5cm,right=2.5cm,top=2.5cm,bottom=2.5cm,letterpaper]{geometry}
\usepackage[doublespacing]{setspace}
\usepackage{graphicx}
\usepackage{amssymb}
\usepackage[foot]{amsaddr}
\usepackage{epstopdf}
\usepackage[colorlinks,citecolor=blue,urlcolor=blue,linkcolor=blue,bookmarks=false,hypertexnames=true]{hyperref}

\title[Maximum entropy model and cardiac fibrillation dynamics]{Energy landscape analysis of cardiac fibrillation wave dynamics using pairwise maximum entropy model}

\author[E. Song]{Euijun Song\\
\it{Yonsei University College of Medicine, Seoul, Republic of Korea}}
\email{drjunsong@gmail.com}
\subjclass{92C05}
\keywords{cardiac fibrillation, spiral wave dynamics, maximum entropy model, energy landscape, computational modeling}

\begin{document}
\begingroup
\def\uppercasenonmath#1{} 
\let\MakeUppercase\relax 
\maketitle
\endgroup

\begin{abstract}
Cardiac fibrillation is characterized by chaotic and disintegrated spiral wave dynamics patterns, whereas sinus rhythm shows synchronized excitation patterns. To determine functional correlations among cardiomyocytes during complex fibrillation states, we applied a pairwise maximum entropy model (MEM) to the 2D numerical simulation data of human atrial fibrillation. We then constructed an energy landscape and estimated a hierarchical structure among the different local minima (attractors) to explain the dynamic properties of cardiac fibrillation. The MEM could describe the wave dynamics of sinus rhythm, single stable rotor, and single rotor with wavebreak (both accuracy and reliability$>$0.9), but not the multiple random wavelet case. The energy landscapes exhibited unique profiles of local minima and energy barriers, characterizing the spatiotemporal patterns of cardiac fibrillation dynamics. The pairwise MEM-based energy landscape analysis provides reliable explanations of complex nonlinear dynamics of cardiac fibrillation, which might be determined by the presence of a 'driver' such as a sinus node or rotor.
\end{abstract}

\section{Introduction}

The human heart consists of billions of cardiomyocytes, electrically connected through cell-to-cell gap junctions. The abnormal and chaotic electrical wave propagation on the cardiac tissue causes harmful or life-threatening cardiac arrhythmias, including atrial fibrillation, ventricular tachycardia, and ventricular fibrillation. Two major spatiotemporal patterns of cardiac fibrillation are spiral wave reentry \cite{Gray98, Gray95, Mandapati} and multiple wavelet \cite{Moe62, Moe64}, which independently or simultaneously contribute to the initiation and maintenance of cardiac fibrillation. However, the biophysical mechanism of cardiac fibrillation is still unclear because of the complex nonlinear nature the cardiac system \cite{Garfinkel}. Identifying the dynamic stability of fibrillation states and transitional properties from disordered fibrillation states to ordered states are important to discover novel mechanisms and therapeutic strategies of cardiac arrhythmias.

Ashikaga and colleagues tried to elucidate complex electrical communication between cardiomyocytes perturbed during arrhythmias through the information and network theoretical approaches \cite{Ashikaga15, Ashikaga18a, Ashikaga18b}. Similar approaches have been well established in systems neuroscience fields to study the dynamical organization and disorganization of brain activity \cite{Park, Tononi, Deco}. One feasible method to explain nonlinear dynamics in resting brain activity is estimating hidden functional interactions and constructing an energy landscape by fitting a pairwise maximum entropy model (MEM) to brain activity data \cite{Schneidman, Watanabe13, Watanabe14, Kang, Ashourvan}. Using the pairwise MEM, the brain dynamics could be explained by the combinations of mean activities and pairwise correlations among the brain regions. Indeed, this 'energy landscape' concept in biological systems is firstly proposed by Waddington to illustrate the cellular differentiation process in terms of epigenetics \cite{Waddington}. The MEM approach provides reliable explanations of diverse complex biological systems, including the immune repertoire diversity \cite{Mora, Arora}, protein signaling networks \cite{Locasale}, and protein folding \cite{Steinbach}.

The pairwise MEM maximizes the entropy $$S=-\sum_{X}p(X)\ln{p(X)}$$ where $X$ is the state of a given dynamical system. The Boltzmann distribution with inverse temperature is usually assumed \cite{Watanabe13, Watanabe14, Kang}, thereby giving the configuration probability $p(X)$ as $p(X) \propto e^{-\beta H}$. The Hamiltonian $H$ of the system is typically estimated from the Ising model, reflecting pairwise correlations of the cells or regions \cite{Stein}.

Since cardiomyocytes are electrically connected similar to the brain, we utilize the pairwise MEM to systemically dissect the spiral wave dynamics of cardiac fibrillation. In the present work, we estimate the energy landscape of cardiac fibrillation dynamics by applying the pairwise MEM to the sequential electrical activity maps. We simulate a two-dimensional computational model of human atrial fibrillation and generate four different types of wave dynamics: sinus rhythm, single stable rotor, single rotor with wavebreak, and multiple wavelet. We finally analyze local minima (i.e., attractors), basin sizes, and energy barriers to describe the wave dynamics of cardiac fibrillation.

\section{Methods}

\subsection{Numerical simulation of cardiac fibrillation}
We simulated a two-dimensional isotropic, homogeneous cardiac tissue ($512\times512$, $\Delta{x}=0.025 ~\rm{cm}$) by numerically solving the following reaction-diffusion equation \cite{Keener}:
\begin{eqnarray} \label{eq:rde}
\frac{\partial V}{\partial t}=-\frac{I_{ion}+I_{stim}}{C_m}+D\nabla^2V
\end{eqnarray}
where $V$ is the transmembrane potential, $I_{ion}$ is the total ionic currents, $I_{stim}$ is the stimulus current, $D=0.001~\rm{cm^{2}/ms}$ is the diffusion coefficient \cite{Pandit, Xie}, and $C_{m}=100$ pF is the membrane capacitance. Each cell in the $512\times512$ grid has an action potential (i.e., the first part of \autoref{eq:rde}) and the cells are electrically connected with adjacent cells (i.e., the last part of \autoref{eq:rde}). We used the Courtemanche-Ramirez-Nattel action potential model of the human atrial cardiomyocyte \cite{Courtemanche}, incorporating 12 transmembrane currents and intracellular calcium handling with a two-compartment sarcoplasmic reticulum. Each ionic current $I_{X}~(X=Na^{+}, K^{+}, Ca^{2+}, etc.)$ is typically a variation of the Hodgkin-Huxley-type model \cite{Hodgkin}, that is, it can be expressed as a nonlinear function of the action potential $V$ and ion channel gating variables $u_i$ as follows:
\begin{eqnarray}
I_{X} &=& G_{X}\prod_{i}{u_{i}(V-E_{m,X})} \nonumber\\
\frac{d u_i}{d t} &=& \alpha_{i}(1-u_{i})-\beta_{i}u_{i} \nonumber
\end{eqnarray}
where $G_{X}$ is the conductance, $E_{m,X}$ is the Nernst potential of the ion $X$, and $\alpha_i, \beta_i$ are the rate constants fitted from experimental voltage-clamp data.
The biophysical details are described in Courtemanche et al. \cite{Courtemanche} and the cell model is available at the CellML Physiome Project (https://models.physiomeproject.org).

Four types of the electrical wave dynamics were considered \cite{Lee}: 1) sinus rhythm (no electrical remodeling, pacing cycle length=500 ms); 2) single stable rotor ($I_{CaL}\times0.3$); 3) single rotor with wavebreak ($I_{Na}\times0.9,~I_{to}\times0.3,~I_{CaL}\times0.5,~I_{K1}\times2,~I_{Kur}\times0.5,~I_{NCX}\times1.4,~I_{leak}\times0.8$); and 4) multiple wavelet ($I_{CaL}\times0.5$). The physiological meanings of the ionic currents are described in \autoref{table:first}. We applied the standard S1S2 cross-field protocol to initiate a spiral wave \cite{Deo13}. An adaptive finite difference method was used to numerically solve the partial differential equation (\autoref{eq:rde}). The adaptive time step of 0.01–0.1 ms was used and the data sampling interval was 1 ms. The numerical simulation was performed using a C++ code with GPU parallelization.

\begin{table}[h!]
\caption{Descriptions of ionic currents in the human atrial cell model.}
\label{table:first}
\begin{tabular}{ll}
\hline
Ionic current & Description \\ \hline
$I_{CaL}$ & L-type calcium current \\
$I_{Na}$ & Fast sodium current \\
$I_{to}$ & Transient outward potassium current \\
$I_{K1}$ & Inward rectifier potassium current \\
$I_{Kur}$ & Ultrarapid delayed rectifier potassium current \\
$I_{NCX}$ & Sodium-calcium exchange current \\
$I_{leak}$ & Calcium leak from sarcoplasmic reticulum to myoplasm \\ \hline
\end{tabular}
\end{table}

\subsection{Pairwise maximum entropy model}
\begin{figure}[h!]
  \includegraphics[width=0.9\linewidth]{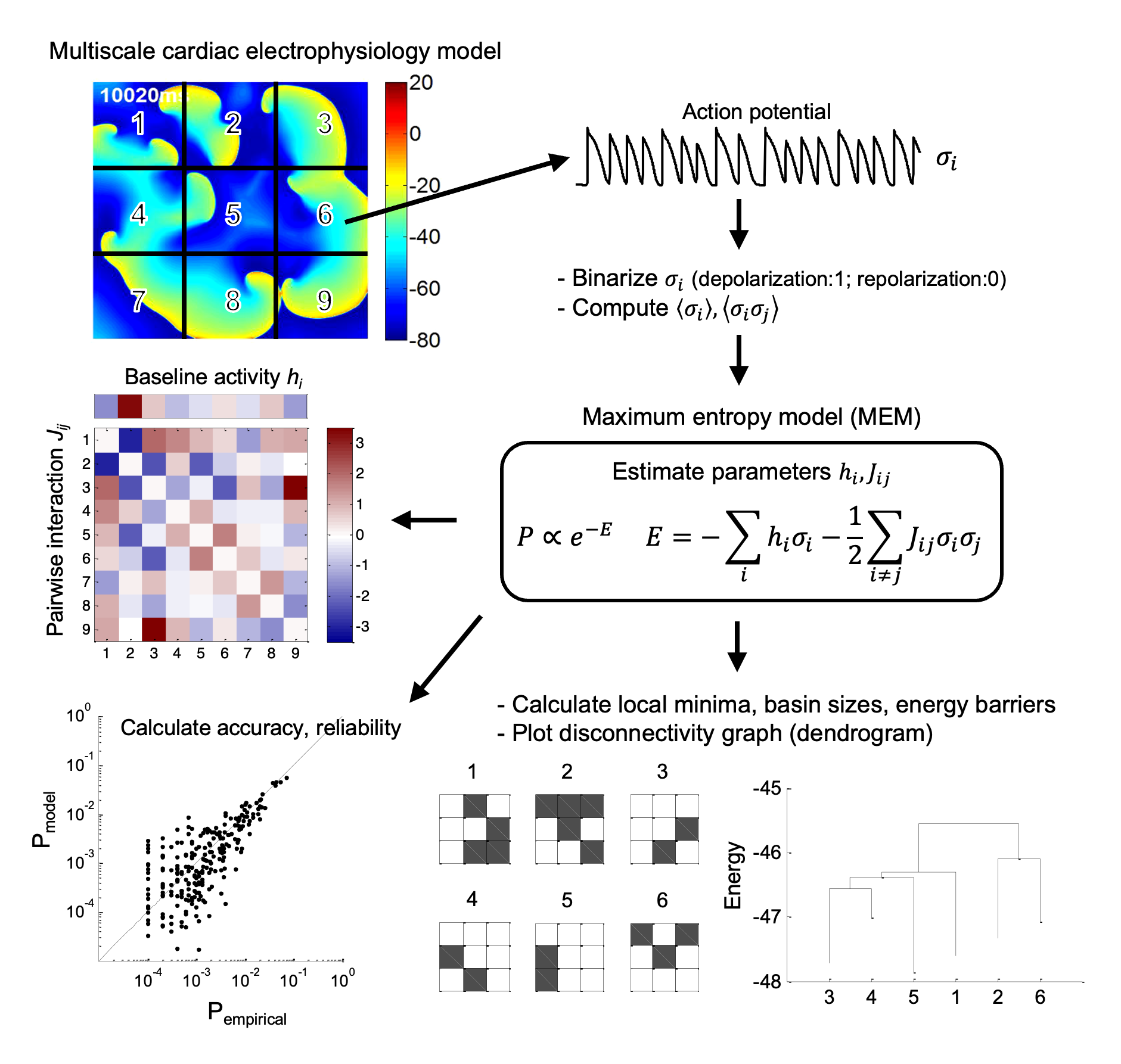}
  \caption{Schema of the energy landscape analysis using a pairwise maximum entropy model (MEM).}
  \label{fig:first}
\end{figure}

We used the pairwise MEM framework as described in Watanabe et al. \cite{Watanabe13, Watanabe14} and Kang et al \cite{Kang}. We fit the pairwise MEM to the simulated electrical activity data of atrial fibrillation (see \autoref{fig:first}). The wave dynamics at time $t$ ($0\leq t\leq T=10$ sec) is represented as vector $V(t)=[\sigma_1,\sigma_2,\ldots,\sigma_N] \in W$ where $N=9$ is the number of the selected action potential data, $\sigma_i$ is the binarized action potential signal, and $W$ is the $N$-dimensional binary state space $W=\{0,1\}^{N}$. We calculated the empirical activation rate $\langle\sigma_i \rangle=\frac{1}{T} \sum_{t=1}^{T}\sigma_{i}^{t}$ and the empirical pairwise joint activation rate $\langle\sigma_i \sigma_j\rangle=\frac{1}{T} \sum_{t=1}^{T}\sigma_{i}^{t} \sigma_{j}^{t}$. Under the constraints preserving $\langle\sigma_i\rangle$ and $\langle\sigma_i \sigma_j\rangle$, the pairwise MEM maximizes the entropy $S=-\sum_{k}p(V_k)\ln{p(V_k)}$, deriving the probability of the state $V_k$ as
\begin{eqnarray}
p(V_k)=\frac{e^{-E(V_k)}}{\sum_{w\in W}{e^{-E(w)}}}.
\end{eqnarray}
Here, we assumed the Boltzmann distribution with inverse temperature and also assumed $\beta=1$ for convenience as previously described \cite{Watanabe13, Watanabe14, Kang}. $E(V_k)$ represents the energy (i.e., Hamiltonian) of the state $V_k$, described as the following Ising model \cite{Landau}:
\begin{eqnarray} \label{eq:ising}
E=-\sum_{i}{h_i\sigma_i} - \frac{1}{2}\sum_{i\neq j}{J_{ij}\sigma_i \sigma_j}
\end{eqnarray}
where $h_i$ represents the tendency of activation at region $i$ (baseline activity) and $J_{ij}$ represents functional interaction between region $i$ and $j$ (pairwise interaction) ($1\leq i,j \leq N=9$). We calculated the expected activation rate $\langle\sigma_i\rangle_m=\sum_{k=1}^{2^N}{\sigma_i (V_k) P(V_k)}$ and the expected pairwise joint activation rate $\langle\sigma_i\sigma_j\rangle_m=\sum_{k=1}^{2^N}{\sigma_i (V_k)\sigma_j (V_k) P(V_k)}$. We used an iterative scaling method to estimate the MEM parameters $h_i$ and $J_{ij}$ as follows:
\begin{eqnarray}
h_i^{new} &=& h_i^{old}+\alpha\ln \frac{\langle\sigma_i\rangle}{\langle\sigma_i\rangle_m} \nonumber\\
J_{ij}^{new} &=& J_{ij}^{old}+\alpha\ln\frac{\langle\sigma_i\sigma_j\rangle}{\langle\sigma_i\sigma_j\rangle_m} \nonumber
\end{eqnarray}
where $\alpha=0.1$ is the learning rate. The accuracy of fit was calculated as $$\text{Accuracy}~r_D=\frac{D_1-D_2}{D_1}$$ where $D_k~(k=1,2)$ is the Kullback-Leibler divergence between the probability distribution of state in $k$th-order model ($P_k$) and the empirical distribution of state ($P_N$) as $D_{k}=D_{\rm{KL}}(P_{N}\parallel P_{k})$. The reliability was calculated as $$\text{Reliability}=\frac{(S_1-S_2)/(S_1-S_N)}{r_D}$$ where $S_k~(k=1,2)$ is the entropy of the distribution of state in $k$th-order model. Here, the second-order model indicates the Ising model (\autoref{eq:ising}), and the energy of the first-order model is defined as $E=-\sum_{i}{h_i\sigma_i}$, ignoring the second-order term in \autoref{eq:ising}.

\subsection{Energy landscape analysis}
We calculated the energy landscape and evaluated the local minima, basin sizes, and energy barriers as described in previous studies \cite{Watanabe14, Kang, Ashourvan}. We first searched the local minima (i.e., attractors) which have lower energy than $N$ adjacent states. All states were classified into the basin of the local minimum by continuously moving the node to the neighbor node with the smallest energy value until reached to the attractor. The basin size was defined by the fraction of states that belonged to the basin of the attractor. We then performed the disconnectivity graph analysis \cite{Becker, Wales} to generate the energy landscape. The energy barrier between two local minima $i$ and $j$ was defined as $$\text{Energy barrier}=\min\{E^{b}(V_i,V_j )-V_i,~E^{b}(V_i,V_j )-V_j\}$$ where $E^{b}(V_i,V_j )$ is the threshold energy level calculated as the highest energy on the shortest path connecting the two local minima. The MATLAB was used for data processing and visualization.

\section{Results and Discussion}

\begin{figure}[h!]
  \includegraphics[width=\linewidth]{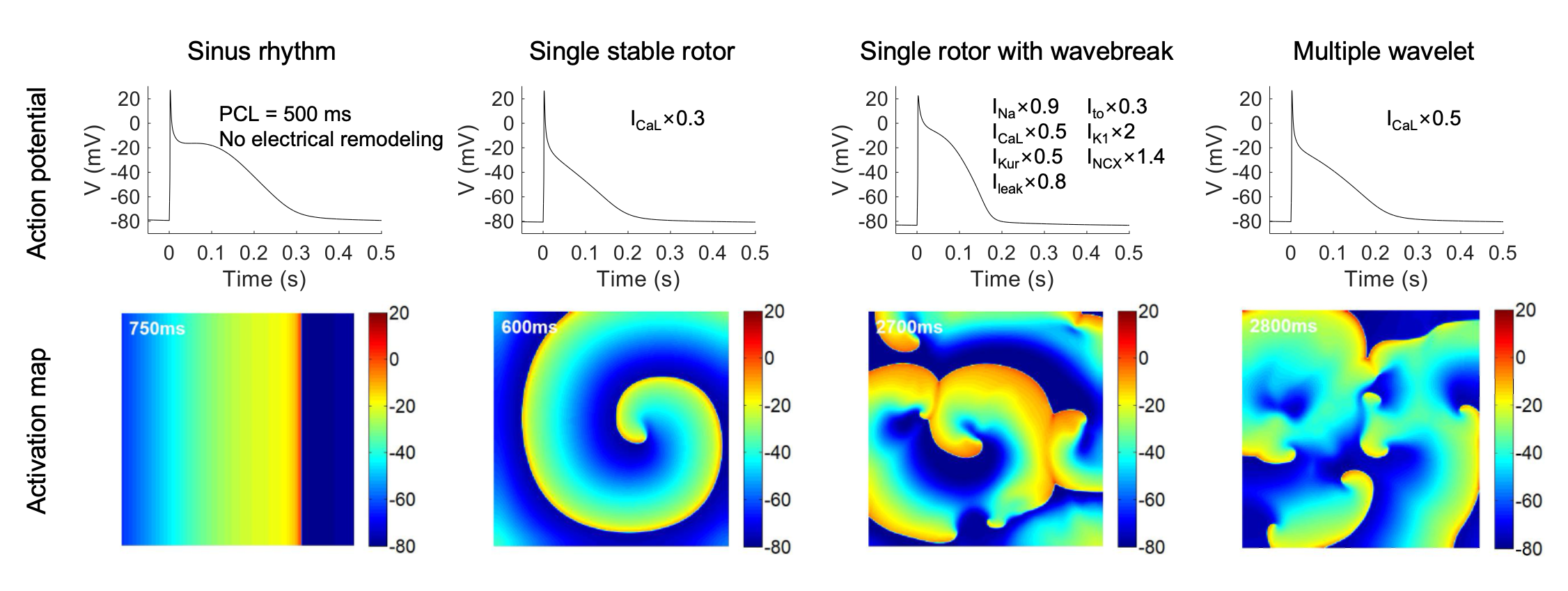}
  \caption{Four types of cardiac wave dynamics: sinus rhythm, single stable rotor, single rotor with wavebreak, and multiple wavelet. The simulated action potentials and two-dimensional activation maps are shown.}
  \label{fig:second}
\end{figure}

We simulated four types of cardiac wave dynamics by perturbing ion channel parameters: sinus rhythm, single stable rotor, single rotor with wavebreak, and multiple wavelet (\autoref{fig:second}). All the fibrillation states were induced by the S1S2 cross-field protocol and were successfully maintained for more than 10 seconds.

\begin{figure}[h!]
  \includegraphics[width=\linewidth]{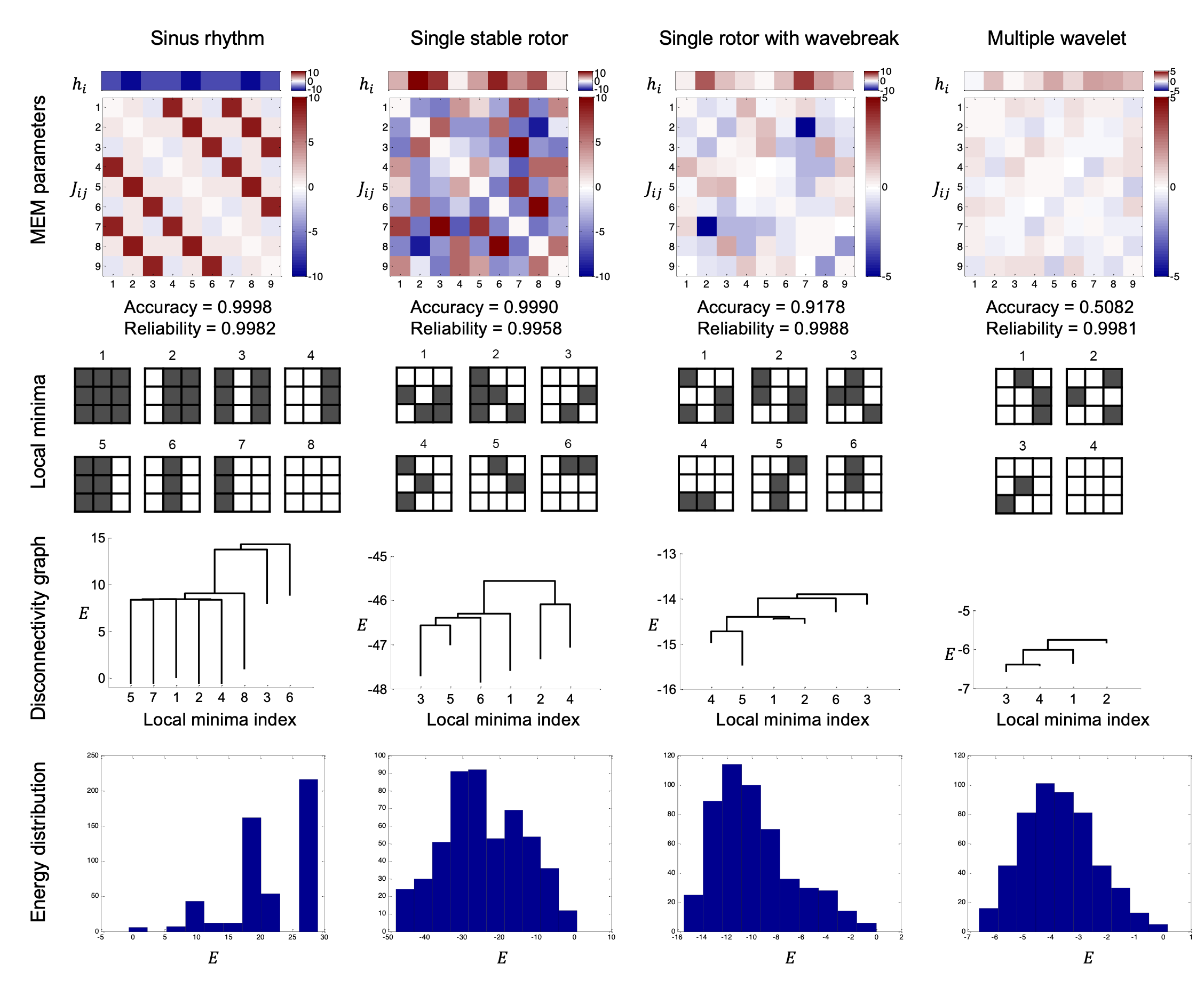}
  \caption{Energy landscape analyses for sinus rhythm and cardiac fibrillation wave dynamics. The coefficients of the pairwise maximum entropy model (MEM), local energy minima (attractors), and energy landscape (disconnectivity graph) are shown.}
  \label{fig:third}
\end{figure}

We then performed the energy landscape analysis by applying the pairwise MEM to the sequential two-dimensional electrical activity data (see \autoref{fig:third}). For all types of wave dynamics except the multiple random wavelet, the pairwise MEMs were fitted to the empirical simulation data with high accuracy ($>$0.90) and high reliability ($>$0.99). However, the multiple wavelet case showed low accuracy ($=$0.5082) and high reliability ($>$0.99). Both of the sinus rhythm and the single stable rotor showed relatively high pairwise interaction coefficients among the cardiomyocytes. In addition, the local energy minima had relatively large basins and high energy barrier, indicating stable attractor properties. This result implies the ‘driving’ role of a sinus node or rotor in the cardiac electrical wave dynamics. However, in the single rotor with wavebreak, there were relatively low pairwise interaction coefficients and a similar number of the local minima separated by a relatively low energy barrier, compared with the single stable rotor case. The energy landscape of the multiple wavelet consisted of a large number of the local minima separated by a relatively low energy barrier, indicating unstable dynamics. Overall results imply that the presence of 'driver' such as a sinus node or rotor plays a pivotal role in determining the dynamical stability of cardiac fibrillation.

Our results demonstrate that our MEM framework can dissect the hidden patterns underlying cardiac wave dynamics. One popular method for evaluating hidden correlation structures underlying nonlinear dynamics of complex biological systems is an information-theoretic approach such as information flow or transport \cite{Ashikaga15, Ashikaga17, Rabinovich}. However, information measures are static and univariate quantities, and thus cannot fully capture the global spatiotemporal dynamics of cardiac fibrillation. Since our MEM framework has broad generalizability, it may allow one to uncover global and local dynamical patterns from clinical intracardiac electrogram signal data.

Although we only considered isotropic and homogeneous cardiac models, our framework can be utilized to explore cardiac fibrillation wave dynamics under heterogeneous electrical/structural remodeling conditions \cite{Bub}. More generally and theoretically, our approaches might be used for any spiral wave or pattern formation dynamics on the excitable media \cite{Barkley90, Barkley94}, as well as patient-specific 3D heart models \cite{Trayanova} and clinically-acquired electroanatomical mapping data \cite{Ganesan}. Since the conventional MEM only captures pairwise interactions and correlations, further studies should be needed to elucidate scaling properties and higher-order correlations to discover hidden complex patterns in cardiac spiral wave dynamics.

\newpage
\section{Conclusions}
The pairwise maximum entropy model revealed the local and global correlations among the cardiomyocytes beyond the simple anatomical connectivity. The energy landscape analysis could explain the stability and transitional properties of complex chaotic dynamics of cardiac fibrillation, which might be determined by the presence of 'driver' such as a sinus node or rotor.

\section*{Acknowledgements}
The pilot result of this study was presented at the \textit{62$^{nd}$ Biophysical Society Annual Meeting}, San Francisco, CA, 2018 (late poster). The author would like to thank T. G. Song and H.-N. Pak for their kind support and anonymous reviewers for their constructive comments.

This research received no external funding. The author has declared no competing interest.

\section*{Author Contributions}
\textbf{E. Song:} Conceptualization, Methodology, Formal analysis, Software, Investigation, Visualization, Writing – original draft, Writing – review \& editing



\begin{thebibliography}{999}
\bibitem{Gray98} Gray, R. A., Pertsov, A. M. \& Jalife, J. Spatial and temporal organization during cardiac fibrillation. Nature 392, 75-78, doi:10.1038/32164 (1998).

\bibitem{Gray95} Gray, R. A. et al. Mechanisms of cardiac fibrillation. Science 270, 1222-1223; author reply 1224-1225 (1995).

\bibitem{Mandapati} Mandapati, R., Skanes, A., Chen, J., Berenfeld, O. \& Jalife, J. Stable microreentrant sources as a mechanism of atrial fibrillation in the isolated sheep heart. Circulation 101, 194-199 (2000).

\bibitem{Moe62} Moe, G. K. On the multiple wavelet hypothesis of atrial fibrillation. Arch Int Pharmacodyn Ther 140, 183 (1962).

\bibitem{Moe64} Moe, G. K., Rheinboldt, W. C. \& Abildskov, J. A. A Computer Model of Atrial Fibrillation. Am Heart J 67, 200-220 (1964).

\bibitem{Garfinkel} Garfinkel, A. et al. Quasiperiodicity and chaos in cardiac fibrillation. The Journal of clinical investigation 99, 305-314, doi:10.1172/jci119159 (1997).

\bibitem{Ashikaga15} Ashikaga, H. et al. Modelling the heart as a communication system. J R Soc Interface 12, doi:10.1098/rsif.2014.1201 (2015).

\bibitem{Ashikaga18a} Ashikaga, H. \& James, R. G. Inter-scale information flow as a surrogate for downward causation that maintains spiral waves. Chaos 28, 075306, doi:10.1063/1.5017534 (2018).

\bibitem{Ashikaga18b} Ashikaga, H. \& Asgari-Targhi, A. Locating Order-Disorder Phase Transition in a Cardiac System. Sci Rep 8, 1967, doi:10.1038/s41598-018-20109-6 (2018).

\bibitem{Park} Park, H. J. \& Friston, K. Structural and functional brain networks: from connections to cognition. Science 342, 1238411, doi:10.1126/science.1238411 (2013).

\bibitem{Tononi} Tononi, G., Edelman, G. M. \& Sporns, O. Complexity and coherency: integrating information in the brain. Trends in Cognitive Sciences 2, 474-484, doi:10.1016/S1364-6613(98)01259-5 (1998).

\bibitem{Deco} Deco, G., Jirsa, V. K. \& McIntosh, A. R. Emerging concepts for the dynamical organization of restingstate activity in the brain. Nature reviews. Neuroscience 12, 43-56, doi:10.1038/nrn2961 (2011).

\bibitem{Waddington} Waddington, C. H. The Strategy of the Genes: A Discussion of Some Aspects of Theoretical Biology. (Allen \& Unwin, London, 1957).

\bibitem{Schneidman} Schneidman, E., Berry, M. J., 2nd, Segev, R. \& Bialek, W. Weak pairwise correlations imply strongly correlated network states in a neural population. Nature 440, 1007-1012, doi:10.1038/nature04701 (2006).

\bibitem{Watanabe13} Watanabe, T. et al. A pairwise maximum entropy model accurately describes resting-state human brain networks. Nat Commun 4, 1370, doi:10.1038/ncomms2388 (2013).

\bibitem{Watanabe14} Watanabe, T., Masuda, N., Megumi, F., Kanai, R. \& Rees, G. Energy landscape and dynamics of brain activity during human bistable perception. Nat Commun 5, 4765, doi:10.1038/ncomms5765 (2014).

\bibitem{Kang} Kang, J., Pae, C. \& Park, H. J. Energy landscape analysis of the subcortical brain network unravels system properties beneath resting state dynamics. Neuroimage 149, 153-164, doi:10.1016/j.neuroimage.2017.01.075 (2017).

\bibitem{Ashourvan} Ashourvan, A., Gu, S., Mattar, M. G., Vettel, J. M. \& Bassett, D. S. The energy landscape underpinning module dynamics in the human brain connectome. Neuroimage 157, 364-380, doi:10.1016/j.neuroimage.2017.05.067 (2017).

\bibitem{Stein} Stein, R. R., Marks, D. S., \& Sander, C. Inferring pairwise interactions from biological data using maximum-entropy probability models. PLoS computational biology 11, e1004182, doi:10.1371/journal.pcbi.1004182 (2015).

\bibitem{Mora} Mora, T., Walczak, A. M., Bialek, W., \& Callan Jr, C. G. Maximum entropy models for antibody diversity. Proceedings of the National Academy of Sciences 107, 5405-5410, doi:10.1073/pnas.1001705107 (2010).

\bibitem{Arora} Arora, R., Kaplinsky, J., Li, A., \& Arnaout, R. Repertoire-based diagnostics using statistical biophysics. bioRxiv 519108, doi: 10.1101/519108 (2019).

\bibitem{Locasale} Locasale, J. W., \& Wolf-Yadlin, A. Maximum entropy reconstructions of dynamic signaling networks from quantitative proteomics data. PloS one 4, e6522, doi:10.1371/journal.pone.0006522 (2009).

\bibitem{Steinbach} Steinbach, P. J., Ionescu, R., \& Matthews, C. R. Analysis of kinetics using a hybrid maximum-entropy/nonlinear-least-squares method: application to protein folding. Biophysical Journal 82, 2244-2255, doi:10.1016/S0006-3495(02)75570-7 (2002).

\bibitem{Keener} Keener, J. P. \& Sneyd, J. Mathematical physiology. 2nd edn, (Springer, 2009).

\bibitem{Pandit} Pandit, S. V. et al. Ionic determinants of functional reentry in a 2-D model of human atrial cells during simulated chronic atrial fibrillation. Biophysical Journal 88, 3806-3821, doi:10.1529/biophysj.105.060459 (2005).

\bibitem{Xie} Xie, F., Qu, Z., Garfinkel, A. \& Weiss, J. N. Electrical refractory period restitution and spiral wave reentry in simulated cardiac tissue. Am J Physiol Heart Circ Physiol 283, H448-460, doi:10.1152/ajpheart.00898.2001 (2002).

\bibitem{Courtemanche} Courtemanche, M., Ramirez, R. J. \& Nattel, S. Ionic mechanisms underlying human atrial action potential properties: insights from a mathematical model. Am J Physiol 275, H301-321, doi:10.1152/ajpheart.1998.275.1.H301 (1998).

\bibitem{Hodgkin} Hodgkin, A. L. \& Huxley, A. F. A quantitative description of membrane current and its application to conduction and excitation in nerve. Journal of Physiology 117(4), 500-44, doi:10.1113/jphysiol.1952.sp004764 (1952).

\bibitem{Lee} Lee, Y. S. et al. The Contribution of Ionic Currents to Rate-Dependent Action Potential Duration and Pattern of Reentry in a Mathematical Model of Human Atrial Fibrillation. PLoS One 11, e0150779, doi:10.1371/journal.pone.0150779 (2016).

\bibitem{Deo13} Deo, M. et al. KCNJ2 mutation in short QT syndrome 3 results in atrial fibrillation and ventricular proarrhythmia. Proceedings of the National Academy of Sciences 110(11), 4291-4296, doi:10.1073/pnas.1218154110 (2013).

\bibitem{Landau} Landau, L. D. \& Lifshitz, E. M. Statistical physics. 3rd edn, (Butterworth-Heinemann, 1980).

\bibitem{Becker} Becker, O. M. \& Karplus, M. The topology of multidimensional potential energy surfaces: Theory and application to peptide structure and kinetics. The Journal of Chemical Physics 106, 1495-1517, doi:10.1063/1.473299 (1997).

\bibitem{Wales} Wales, D. J., Miller, M. A., \& Walsh, T. R. Archetypal energy landscapes. Nature 394(6695), 758-760, doi:10.1038/29487 (1998).

\bibitem{Ashikaga17} Ashikaga, H. \& James, R. G. Hidden structures of information transport underlying spiral wave dynamics. Chaos 27(1), 013106, doi:10.1063/1.4973542 (2017).

\bibitem{Rabinovich} Rabinovich, M. I., Afraimovich, V. S., Bick, C., \& Varona, P. Information flow dynamics in the brain. Physics of life reviews 9(1), 51-73, doi:10.1016/j.plrev.2011.11.002 (2012).

\bibitem{Bub} Bub, G., Shrier, A., \& Glass, L. Spiral wave generation in heterogeneous excitable media. Physical review letters 88(5), 058101, doi:10.1103/PhysRevLett.88.058101 (2002).

\bibitem{Barkley90} Barkley, D., Kness, M., \& Tuckerman, L. S. Spiral-wave dynamics in a simple model of excitable media: The transition from simple to compound rotation. Physical Review A 42(4), 2489, doi:10.1103/PhysRevA.42.2489 (1990).

\bibitem{Barkley94} Barkley, D. \& Kevrekidis, I. G. A dynamical systems approach to spiral wave dynamics. Chaos 4(3), 453-460, doi:10.1063/1.166023 (1994).

\bibitem{Trayanova} Trayanova, N. A. Whole-heart modeling: applications to cardiac electrophysiology and electromechanics. Circulation research 108(1), 113-128, doi:10.1161/CIRCRESAHA.110.223610 (2011).

\bibitem{Ganesan} Ganesan, A. N. et al. Bipolar electrogram shannon entropy at sites of rotational activation: implications for ablation of atrial fibrillation. Circulation: Arrhythmia and Electrophysiology 6(1), 48-57, doi:10.1161/CIRCEP.112.976654 (2013).

\end{thebibliography}
\end{document}